\documentclass[10pt,twocolumn,preprintnumbers,amsmath,amssymb,nofootinbib
,superscriptaddress]{revtex4-1}
\usepackage{graphicx,longtable,mathrsfs,color,array}
\usepackage[hidelinks]{hyperref}
\usepackage[usenames,dvipsnames]{xcolor} % colour
\usepackage{amssymb,amsmath,mathtools,mathrsfs,slashed} % maths
\usepackage{epsfig,subfigure,placeins,float} % plots
\usepackage{booktabs,longtable,ctable,multirow} % tables
\usepackage{exscale,relsize} % scale 
\usepackage[normalem]{ulem} % editing
\usepackage{enumerate}
\usepackage{times, mathptmx} % Times font
\usepackage[utf8]{inputenc}
\usepackage{color}
\usepackage{hyperref}
\usepackage{graphicx}% Include figure files
\usepackage{color}
\usepackage{graphicx,graphics}
\usepackage{bm}
\usepackage{tabularx}
\allowdisplaybreaks[1]

\begin{document}

\title{Forecasts for Low Spin Black Hole Spectroscopy in Horndeski Gravity}
\author{Oliver J. Tattersall}
\email{oliver.tattersall@physics.ox.ac.uk}
\affiliation{Astrophysics, University of Oxford, DWB, Keble Road, Oxford OX1 3RH, UK}
\author{Pedro G. Ferreira}
\email{p.ferreira1@physics.ox.ac.uk}
\affiliation{Astrophysics, University of Oxford, DWB, Keble Road, Oxford OX1 3RH, UK}

\date{Received \today; published -- 00, 0000}

\begin{abstract}
We investigate the prospect of using black hole spectroscopy to constrain the parameters of Horndeski gravity through observations of gravitational waves from perturbed black holes. We study the gravitational waves emitted during ringdown from black holes without hair in Horndeski gravity, demonstrating the qualitative differences between such emission in General Relativity and Horndeski theory. In particular, Quasi-Normal Mode frequencies associated with the scalar field spectrum can appear in the emitted gravitational radiation. Analytic expressions for error estimates for both the black hole and Horndeski parameters are calculated using a Fisher Matrix approach, with constraints on the `effective mass' of the Horndeski scalar field of order $\sim 10^{-17}$eV$c^{-2}$ or tighter being shown to be achievable in some scenarios. Estimates for the minimum signal-noise-ratio required to observe such a signal are also presented.
\end{abstract}

\maketitle
%%%%%%%%%%%%%%%%%%%%%%%%%%%%%%%%%%%%%%%%%%%%%%%%%%%%%%%%%%%%%%
\section{Introduction}

The advent of gravitational wave (GW) astronomy, with numerous observations of mergers of compact objects now made by advanced LIGO and VIRGO  \cite{LIGOScientific:2018mvr}, has opened new avenues for testing Einstein's theory of General Relativity (GR) \cite{Einstein:1916vd,Barack:2018yly}. With these tests have come constraints on the landscape of modified gravity theories, most significantly those constraints garnered from the propagation of gravitational waves over cosmological distances \cite{Lombriser:2015sxa,Lombriser:2016yzn,2017PhRvL.119y1301B,Creminelli:2017sry,Sakstein:2017xjx,Ezquiaga:2017ekz}. In the future, next generation ground and space based GW detectors bring the promise of black hole spectroscopy: observing the frequency spectrum of gravitational waves emitted by perturbed black holes -- specifically the Quasi-Normal Mode (QNM) spectrum -- and using them as a fingerprint to both infer the properties of the emitting black holes, as well as to test the predictions of GR against competing theories of gravity \cite{Dreyer:2003bv,Berti:2005ys,Gossan:2011ha,Meidam:2014jpa,Berti:2015itd,Berti:2016lat,Berti:2018vdi,Baibhav:2018rfk}. 

When considering testing the predictions of GW emission between various theories of gravity, one can look for discrepancies induced either by: (a) a background black hole solution in a modified theory of gravity that differs from the standard GR description of black holes (i.e. the Kerr metric for realistic astrophysical sources) and/or by (b) a different dynamical evolution of gravitational waves with respect to GR from the perturbed system, regardless of the properties of the background black hole. Most focus in current research is on (a), looking for shifts in the GR QNM spectrum; this has lead to a flurry of work on the generalised scalar-Gauss-Bonnet theory \cite{Blazquez-Salcedo:2016enn,Blazquez-Salcedo:2017txk,Silva:2017uqg,Antoniou:2017acq,Antoniou:2017hxj,Bakopoulos:2018nui,Silva:2018qhn,Minamitsuji:2018xde,Sullivan:2019vyi,Ripley:2019irj,Macedo:2019sem}, among others \cite{Konoplya:2001ji,Dong:2017toi,Endlich:2017tqa,Cardoso:2018ptl,Brito:2018hjh,Franciolini:2018uyq,Cardoso:2019mqo}. The emphasis in this work will be very different and on (b): we will focus on the case, where the background black hole solution is given by the same solution as in GR, but the emitted GW signal is governed by modified equations of motion leading to {\it extra} QNMs. 

The motivation for pursuing this line of research comes from \cite{PhysRevD.97.084005} where we argued that scalar-tensor theories in which the scalar field had cosmological relevance would, in general, have black holes with no-hair. But more generally, case (b) can be considered a more generic prediction of extensions to GR: while black holes might or might not have hair, the field equations will, for sure, be modified leading to extra QNMs \cite{Barausse:2008xv,Molina:2010fb,2018PhRvD..97d4021T,Witek:2018dmd}.

In this paper we will consider the Horndeski family of scalar-tensor theories of gravity \cite{Horndeski:1974wa}, following on from work in \cite{2018PhRvD..97d4021T,Tattersall:2018nve}. We will first demonstrate the qualitative differences in GW emission in GR and Horndeski gravity -- the presence of extra QNMs -- before quantifying what one could learn about the parameters of a GW emitting system given an observation of sufficient `loudness'. We will show that, in certain circumstances, the observation of (at least) two distinct frequencies of the damped gravitational waves emitted during `ringdown' from a black hole in Horndeski gravity can provide strong constraints on fundamental parameters of the theory. 

\textit{Summary}: In section \ref{secbackground} we will introduce the action for Horndeski gravity and look at black hole perturbations in this family of theories. We will then move on to parameter estimation in section \ref{secparamest}; we introduce a simple analytical model for the gravitational waves emitted from a perturbed black hole in Horndeski gravity and, building on \cite{Berti:2005ys}, we employ a Fisher Matrix analysis to predict quantitatively what one could learn about the fundamental theory from a GW observation. This will include both analytic and numerical results, as well as considerations on the properties of the observed GW signal required to perform such an analysis. In section \ref{discussion} we will discuss the results and limitations of our analysis before making some concluding remarks.

Throughout we will use natural units with $G=c=1$, except where otherwise stated.

\section{Horndeski Gravity}\label{secbackground}

Extensions to General Relativity normally involve additional fields (other than the metric and standard matter fields) that interact non-trivially with gravity. In this paper we will focus on the family of Horndeski theories which introduces an additional scalar field.

\subsection{Action}

A general action for scalar-tensor gravity with 2$^{nd}$ order-derivative equations of motion is given by the Horndeski action \cite{Horndeski:1974wa,Kobayashi:2011nu}:
\begin{align}
S=\int d^4x\sqrt{-g}\sum_{n=2}^5L_n,\label{Shorndeski}
\end{align}
where the Horndeski Lagrangians are given by:
\begin{align}
L_2&=G_2(\phi,X)\nonumber\\
L_3&=-G_3(\phi,X)\Box \phi\nonumber\\
L_4&=G_4(\phi,X)R+G_{4X}(\phi,X)((\Box\phi)^2-\phi^{\alpha\beta}\phi_{\alpha\beta} )\nonumber\\
L_5&=G_5(\phi,X)G_{\alpha\beta}\phi^{\alpha\beta}-\frac{1}{6}G_{5X}(\phi,X)((\Box\phi)^3 \nonumber\\
& -3\phi^{\alpha\beta}\phi_{\alpha\beta}\Box\phi +2 \phi_{\alpha\beta}\phi^{\alpha\sigma}\phi^{\beta}_{\sigma}),
\end{align}
where $\phi$ is the scalar field with kinetic term $X=-\phi_\alpha\phi^\alpha/2$, $\phi_\alpha=\nabla_\alpha\phi$, $\phi_{\alpha\beta}=\nabla_\alpha\nabla_\beta\phi$, and $G_{\alpha\beta}=R_{\alpha\beta}-\frac{1}{2}R\,g_{\alpha\beta}$ is the Einstein tensor. The $G_i$ denote arbitrary functions of $\phi$ and $X$, with derivatives $G_{iX}$ with respect to $X$. GR is given by the choice $G_4=M_{P}^2/2$ with all other $G_i$ vanishing and $M_{P}$ being the reduced Planck mass. Note that eq.~(\ref{Shorndeski}) is \textit{not} the most general action for scalar-tensor theories, and it has been shown that it can be extended to an arbitrary number of terms \cite{Zumalacarregui:2013pma,Gleyzes:2014qga,Gleyzes:2014dya,Achour:2016rkg}.

\subsection{Background Solutions and Perturbations}

Let us assume that Horndeski gravity admits a background solution such that the spacetime is Ricci flat, $R_{\alpha\beta}=0=R$ (e.g. Minkowski or Schwarzschild), and the scalar field $\phi$ has a trivial constant profile, $\phi=\phi_0$. Several no-hair theorems for various manifestations of Horndeski gravity, leading to such solutions, exist in the literature \cite{Hui:2012qt,Sotiriou:2015pka,Maselli:2015yva,Sotiriou:2015pka}. 

Now consider perturbations to this background solution such that:
\begin{align}
g_{\alpha\beta}=&\;\bar{g}_{\alpha\beta}+h_{\alpha\beta}\\
\phi=&\;\phi_0+\delta\phi,
\end{align}
where $\delta\phi$ and $h_{\alpha\beta}$ are considered to be small and of the same perturbative order, and the metric $\bar{g}_{\alpha\beta}$ describes the background Ricci-flat spacetime. Varying the action given by eq.~(\ref{Shorndeski}) with respect to $g_{\alpha\beta}$ and $\phi$, we find the following system of equations for the perturbed fields:
\begin{subequations}
\begin{align}
G_{\alpha\beta}^{(1)}=&\;\frac{G_{4\phi}}{G_4}\left(\nabla_\alpha\nabla_\beta \delta\phi - g_{\alpha\beta}\Box \delta\phi\right)\label{einseq}\\
\Box\delta\phi=&\;\mu^2\delta\phi\label{scalareq}
\end{align}
\end{subequations}
where $G_{\alpha\beta}^{(1)}$ is the Einstein tensor perturbed to linear order in $h_{\mu\nu}$, $\Box=\bar{g}^{\mu\nu}\nabla_\mu\nabla_\nu$ and $\mu^2$ is given by:
\begin{align}
\mu^2=&\; \frac{-G_{2\phi\phi}}{G_{2X}-2G_{3\phi}+3G_{4\phi}^2/G_4}.\label{musquared}
\end{align}
All of the Horndeski $G_i$ functions are evaluated at the background i.e. they are functions of the constant $\phi_0$ only. Thus $\mu^2$ is a constant and acts like an effective mass term squared for the scalar field. Eq.~(\ref{musquared}) clearly shows that for some combination of the Horndeski parameters, $\mu^2$ could be negative. We will assume that $\mu^2>0$ for the rest of this work, but considering a negative effective mass squared could be an interesting area of future research. Furthermore, we will assume that $G_{2X}-2G_{3\phi}+3G_{4\phi}^2/G_4\neq0$. The right hand side of eq.~(\ref{einseq}) shows the new gravitational scalar field sourcing the gravitational perturbations. 

\subsection{Black Holes and Ringdown}

The sourcing of gravitational perturbations by the Horndeski scalar field, as shown by eqs.~(\ref{einseq})-(\ref{scalareq}) was discovered and investigated in \cite{2018PhRvD..97d4021T,Tattersall:2018nve}, in the particular case that the background solution was a Schwarzschild black hole. However here we can see that any Ricci flat black hole solution with a constant scalar field profile (i.e. no non-trivial scalar hair) will exhibit the same behaviour as, for example, a Kerr black hole. Also note further that in \cite{Tattersall:2018nve} a second parameter $\Gamma$ was erroneously included in the equation of motion for $\phi$; eq.~(\ref{scalareq}) is in the correct form with $\mu^2$ entirely describing the effect of the various Horndeski functions in the scalar equation of motion. 

For the rest of this paper we will be concerned with the effect of Horndeski gravity on the GW signal emitted from a black hole as it `rings down' following a merger event or some other process which leaves the black hole perturbed. To do so, we can consider the background spacetime to be Schwarzschild and decompose the metric and scalar perturbations into tensor spherical harmonics, as is standard when studying the response of spherically symmetric black holes to perturbations \cite{1975RSPSA.343..289C,0264-9381-16-12-201,Kokkotas:1999bd,Berti:2009kk,Martel:2005ir,Ripley:2017kqg}. We will further assume a harmonic time dependence of $e^{-i\omega t}$ for the metric and scalar perturbations, leading to the following system of equations:
\begin{subequations}
\begin{align}
\frac{d^2Q}{dr_\ast^2}+ \left[\omega^2-V_{RW}(r)\right]Q=&\;0\label{RWeq}\\
\frac{d^2\Psi}{dr_\ast^2}+ \left[\omega^2-V_Z(r)\right]\Psi=&\;\frac{G_{4\phi}}{G_4}S_\varphi\left(\varphi,\varphi'\right)\label{Zeq}\\
\frac{d^2\varphi}{dr_\ast^2}+\left[\omega^2-V_S(r,\mu^2)\right]\varphi=&\;0,\label{Phieq}
\end{align}
\end{subequations}
where $\varphi(r)$ is the radial wave function of $\delta\phi$, and $Q(r)$ and $\Psi(r)$ represent the odd and even parity degrees of freedom in the gravitational perturbations (and $r$ is the usual Schwarzschild radial coordinate). The tortoise coordinate $r_\ast$ is defined by $dr_\ast = (1-2M/r)^{-1}dr$, with $M$ being the black hole mass. 

The various potentials appearing in the above equations are given by:
\begin{subequations}
\begin{align}
V_{RW}(r)=&\;\left(1-\frac{2M}{r}\right)\left(\frac{\ell(\ell+1)}{r^2}-\frac{6M}{r^3}\right)\\
V_Z(r)=&\;2\left(1-\frac{2M}{r}\right)\frac{\lambda^2r^2\left[(\lambda+1)r+3M\right]+9M^2(\lambda r+M)}{r^3(\lambda r+3M)^2}\\
V_S(r)=&\left(1-\frac{2M}{r}\right)\left(\frac{\ell(\ell+1)}{r^2}+\frac{2M}{r^3}+\mu^2\right)
\end{align}
\end{subequations}
with $2\lambda=\;(\ell+2)(\ell-1)$, whilst the `source term' $S_\varphi\left(\varphi,\varphi'\right)$ is given by:
\begin{align}
S_\varphi\left(\varphi,\varphi'\right)=&\;\frac{-(1-2M/r)U_\varphi\left(\varphi,\varphi'\right)}{2r^2\left(\lambda r + 3M\right)^2}\nonumber\\
U_{\varphi}\left(\varphi,\varphi'\right)=&\;\left(4M\left(3M+\left(2\ell(\ell+1)-1\right)r\right)\right.\nonumber\\
&\left.+2r^3\left(3M+\lambda r\right)\mu^2\right)\varphi+12M r (r-2M)\varphi'.
\end{align}
Note that we have suppressed spherical harmonic indices, but each equation is assumed to hold for a given $\ell$. 

These equations can be solved to find the complex solutions of the $\omega$, the QNM frequencies, subject to boundary conditions such that the emitted waves are are purely ingoing at the black hole horizon and purely outgoing at spatial infinity \cite{1975RSPSA.343..289C,0264-9381-16-12-201,Kokkotas:1999bd,Berti:2009kk,0264-9381-16-12-201}. The fact that the $\omega$ are complex means that the gravitational waves emitted from this system are not only oscillatory in nature, but also decay in time: the gravitational waves emitted are essentially exponentially damped sinusoids. The $\omega$ describe the oscillation frequency $f$ and damping time $\tau$ of the gravitational (and, in this case, scalar) waves via
\begin{align}
\omega_{\ell m}=2\pi f_{\ell m} +i/\tau_{\ell m},
\end{align}
where we have reinserted spherical harmonic indices $\ell$ and $m$ to emphasise that these relations are general for any $\omega_{\ell m}$. 

The system of equations given by eq.~(\ref{RWeq})-(\ref{Phieq}) shows that the odd parity gravitational degree of freedom $Q$ is decoupled from $\varphi$ and evolves exactly as in GR according to the Regge-Wheeler equation \cite{Regge:1957td}. On the other hand, whilst the even parity gravitational field $\Psi$ obeys the well known Zerilli equation as in GR \cite{Zerilli:1970se} on the `left hand side' of eq.~(\ref{Zeq}), $\Psi$ is now also sourced by the freely evolving scalar wave function $\varphi$. 

We can interpret eq.~(\ref{Zeq}) and eq.~(\ref{Phieq}) as leading to the gravitational field oscillating with both the `transient' GR solution and the `driving' scalar field solution. An analogous mode-mixing situation arises in a certain parameter limit of Chern-Simons gravity \cite{Molina:2010fb}, where it is the odd parity gravitational degree of freedom that is driven by a free massless scalar field. In this case a two-mode fit of each of the fundamental $\ell=2$ modes from the gravitational and scalar spectra fits the numerical evolution of the QNM equations well. In \cite{Witek:2018dmd} the `reverse' effect was observed numerically in scalar Gauss-Bonnet gravity. In this case, again due to a system of coupled QNM equations, the emitted \textit{scalar} waves appear to be `contaminated' with modes arising from the gravitational spectrum. 

A similar perturbation analysis can be done for a slowly rotating Kerr black hole (leading to more complex, but qualitatively similar equations of motion to eq.~(\ref{RWeq})-(\ref{Phieq})). For perturbations of Kerr black holes of arbitrary spin, however, one requires the Teukolsky equation \cite{1973ApJ...185..635T}. Eq.~(\ref{einseq}) shows that any modified gravity effects on the `right hand side' of the equation can be packaged as the `source' of the Teukolsky equation. This opens up the possibility of analytically studying perturbations to black holes of arbitrary spin in modified gravity, provided the background scalar field has a trivial constant profile. We will leave such an analysis to future work.

It is worth noting that for \textit{massive} scalar fields there exists a second family of solutions other than the QNMs, the `quasi-bound states', that represent long lived field configurations around the black hole \cite{Detweiler:1980uk,Furuhashi:2004jk,Cardoso:2005vk,Dolan:2007mj,Cardoso:2011xi,Hod:2012zza,Hod:2012px,Witek:2012tr}. We will not consider the bound states in this work and instead focus on the QNM family of solutions driving the emitted gravitational waves. A full analysis of the black hole-scalar system for arbitrary spin black holes would, of course, warrant consideration of such bound states. It is an intriguing question as to how the quasi-bound states around black holes might drive gravitational radiation during ringdown.

\begin{figure*}
\caption{Waveform `cartoon' of a superposition of $\ell=2$ gravitational and massless scalar modes for a unit mass Schwarzschild black hole. The amplitude of the gravitational waveform is fixed to $A_g=1$ with $A_s$ varying from $0.1$ to $1$.}
\label{cartoon}
\includegraphics[width=\textwidth]{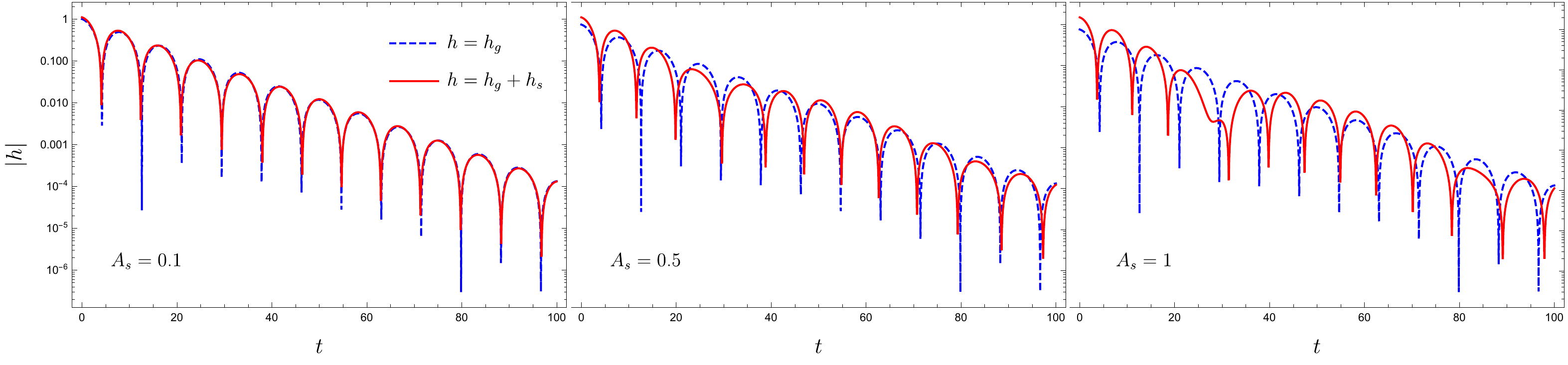}\\
\end{figure*}

Figure \ref{cartoon} shows the qualitative effect of the scalar perturbations `driving' the gravitational waves. The gravitational waves are modulated from their usual GR frequencies by the frequency of the scalar mode. The scalar amplitude in the centre and right hand panels of figure \ref{cartoon} is exaggerated to make the effect noticeable to the eye, but the qualitative picture of the mode mixing effect is valid.  

As has been discussed before, but bears repeating, the system described by eq.~(\ref{RWeq})-(\ref{Phieq}) exhibits clear non-GR behaviour due to the presence of the scalar field perturbation, despite the background solution being identical to GR (i.e. a Schwarzschild black hole with no scalar hair). Thus the detection of modified gravity effects in the ringdown part of a GW signal is \textit{not necessarily} indicative of violations of the no-hair theorem. Stated another way, even black holes in theories that obey no-hair theorems could exhibit non-GR behaviour in their perturbations \cite{Barausse:2008xv}.

\section{Parameter Estimation}\label{secparamest}

Throughout this section we will follow the formalism laid out by Berti, Cardoso, and Will in \cite{Berti:2005ys} (henceforth referred to as BCW). Readers should refer to BCW for a full review; we will recap the main elements here.

\subsection{Fisher Matrix formalism}

We wish to study the response of a gravitational wave detector to the exponentially damped sinusoidal gravitational waves emitted by a perturbed black hole in Horndeski gravity, as described in section \ref{secbackground}. Firstly, we assume that each of the gravitational waveforms received at our detector is such that the strain $h$ is given by:
\begin{align}
h=h_+F_+ + h_\times F_\times \label{basich}
\end{align}
where $h_+$ and $h_\times$ are the two polarisations of the GW given by (in the frequency domain):
\begin{align}
\tilde{h}_+=&\;A_+ \left[S_{\ell m}e^{i\phi^+} b_+ + S_{\ell m}e^{-i\phi^+} b_-\right] \\
\tilde{h}_\times = &\; -i N_\times A_+ \left[S_{\ell m}e^{i\phi^\times} b_+ - S_{\ell m}e^{-i\phi^\times} b_-\right],
\end{align}
with
\begin{align}
b_{\pm}=&\; \frac{1/\tau_{\ell m}}{(1/\tau_{\ell m})^2+4\pi^2(f\pm f_{\ell m})^2}.
\end{align}
and $\phi^\times=\phi^+ + \phi^0$.

The $F_{+,\times}$ pattern functions represent detector orientation and source direction dependence (see eq.~(3.3) in BCW, and appendix \ref{fisherapp}), while the $S_{\ell m}$ are spin-2 spheroidal functions that are in principle complex. As explained in BCW, however, we will assume that $S_{\ell m}\approx \mathfrak{R}(S_{\ell m})$ due to $\mathfrak{R}(S_{\ell m}) \gg \mathfrak{I}(S_{\ell m})$ for slowly damped modes. This will allow us in the following to make use of the angle averages $<S_{\ell m}^2>=1/4\pi$, $<F_+^2>=<F_\times^2>=1/5$, and $<F_+F_\times>=0$. 

Finally, we assume that the gravitational waves emitted from a system governed by eqs.~(\ref{Zeq})-(\ref{Phieq}) are given as a superposition of two modes, the most dominant mode of each of the gravitational and scalar spectra, such that the total strain is given by:
\begin{align}
h=h_g+h_s,\label{htotal}
\end{align}
where $h_g$ and $h_s$ have the same functional form as outlined above, however with different amplitudes, phases, and appropriate frequencies and damping times.

We can define an inner product in frequency space between two generic waveforms $h_1$ and $h_2$ using the noise spectral density of the detector $S_h(f)$:
\begin{equation}
(h_1|h_2)\equiv2\int_0^\infty\frac{\tilde{h}_1^\ast \tilde{h}_2+\tilde{h}_2^\ast \tilde{h}_1}{S_h(f)}df.
\end{equation}
The signal to noise ratio (SNR) $\rho$ of the signal given by eq.~(\ref{htotal}) is thus:
\begin{align}
\rho^2 =  (h|h)=4\int_0^\infty \frac{\tilde{h}^\ast(f)\tilde{h}(f)}{S_h(f)}df,
\end{align}
and the components of the Fisher matrix $\Gamma_{ab}$ associated with the signal are given by:
\begin{align}
\Gamma_{ab}\equiv \left. \left(\frac{\partial h}{\partial \theta^a}\right | \frac{\partial h}{\partial \theta^b}\right),\label{fisher}
\end{align}
where the $\theta^a$ are the parameters upon which the waveform depends. Note that in each of the above integrals in the frequency domain, we are also implicitly angle-averaging over the sky.

Clearly from eq.~(\ref{fisher}) we will need to evaluate derivatives of the waveform with respect to the various parameters $\theta^a$. For example, in BCW, analytic fits to numerical results for Kerr gravitational QNM frequencies are presented as functions of black hole mass $M$ and dimensionless angular momentum $j$ so that derivatives can be performed analytically. In the case of Horndeski gravity, as studied here, of particular interest is the dependence of the waveform on the effective scalar mass (squared) $\mu^2$. 

For simplicity's sake, when evaluating our analytical results numerically in section \ref{secresults}, we will restrict ourselves to considering (at most) slowly rotating Kerr black holes. Clearly we would ideally like to consider black holes with dimensionless spin $\sim 0.7$ to best replicate the events observed by aLIGO/VIRGO so far, but we use the case of a slowly rotating black hole as a starting point. Thus, when evaluating parameter derivatives for the rest of this paper, we will use the analytic expressions for massive scalar and gravitational QNM frequencies of slowly rotating black holes given in \cite{2009CQGra..26v5003D,Tattersall:2018axd}. For example, for $j=0, \ell=2$, we can make use of the following:
\begin{widetext}
\begin{align}
\omega_g^{\ell=2}=2\pi f_g^{\ell=2}+i/\tau_g^{\ell=2}\approx&\;\frac{1}{M}\left(0.374-0.0887i\right)\nonumber\\
\omega_s^{\ell=2}=2\pi f_s^{\ell=2}+i/\tau_s^{\ell=2}\approx&\;\frac{1}{M}\left(0.484+0.316 \, (\mu M)^2+0.0372 \, (\mu M)^4 + 0.0232 \, (\mu M)^6\right.\nonumber\\
&\left. - \left[0.0968 - 0.108 \, (\mu M)^2 - 0.0272 \, (\mu M)^4 - 0.0246 \, (\mu M)^6\right]i\right)
\end{align}
\end{widetext}
We do, however, emphasise that the analytic expressions presented in section \ref{secresults} are applicable to \textit{any black hole} emitting a mixed mode waveform - it is only in the numerical evaluations of these expressions that we have chosen to limit ourselves to at most slowly rotating Kerr black holes. 

We are now in a position to analytically calculate the SNR and Fisher Matrix components for our mixed mode GW signal, from which we can calculate error estimates from the covariance matrix $\Sigma_{ab}=(\Gamma_{ab})^{-1}$. To do so we will use the `$\delta$-function approximation' introduced in BCW to evaluate frequency integrals, replacing products of the $b_{\pm}(f)$ with appropriately normalised $\delta$-functions. In doing so we assume that $S_h(f_{g,\ell m})\approx S_h(f_{s,\ell m}) = S$, which is appropriate given that $f_{g,\ell m}$ and $f_{s,\ell m}$ will be very close to each other in practice. 

From now on we will suppress the $(\ell,m)$ index on, for example, $f$ and $\tau$, with each expression assumed to hold for a specific choice of harmonic indices. We will retain `g' and `s' subscripts to differentiate between those parameters belonging to the gravitational mode and those belonging to the scalar mode in the mixed mode waveform.

\subsection{Results}\label{secresults}

To find error estimates for each of the parameters of the mixed mode waveform we invert the Fisher matrix and take the diagonal components of $\Sigma_{ab}=(\Gamma_{ab})^{-1}$. In practice, the components of $\Gamma_{ab}$ were first calculated in the parameter basis of $(A_g, \phi_{g}^+, f_g, Q_g, A_s, \phi_{s}^+, f_s, Q_s)$, where the quality factor $Q$ of a mode is given by $Q_{\ell m}=\pi f_{\ell m} \tau_{\ell m}$. We then changed basis from $(f_g, Q_g, f_s, Q_s)$ to $(M, j, \mu^2)$ before inverting and extracting the error estimates.

In the $(f,Q)$ parameter basis, those components of the Fisher Matrix $\Gamma_{ab}$ that do not involve mixing of gravitational and scalar parameters are given in Section IV A of BCW. For those components that do mix gravitational and scalar parameters, analytic expressions are provided in a Mathematica notebook \cite{oxwebsite}, as is an expression for the total SNR $\rho^2$. We do not reproduce the expressions here as in most cases they are exceedingly lengthy and unenlightening. 

From now on we will thus work in a simplified regime were we assume that $N_\times$, $\phi_+$, and $\phi_0$ are known for both gravitational and scalar modes. In particular, we follow the conventions of BCW for a mixed mode waveform, assuming that for each mode $N_\times=1$ with the phases given by $\phi_{g}^+=-\pi/2$, $\phi_{g}^0=\phi_s^0=\pi/2$, $\phi_s^+=-\pi/2+\phi$. With this choice of parameters for the waveform we will be able to work with more digestible analytic expressions.

We will further split our analysis in two separate cases: firstly, fixing $j=0$ leaving us with 4 unknown parameters $(A_g, M, A_s, \mu^2)$ to find error estimates for, and secondly, allowing $j$ to be free (but still constrained to be small, $j\ll1$). 

\subsubsection{$j=0$}

In the case of a Schwarzschild black hole, for a mixed mode waveform with the above choice of parameters for $N_\times$ and phases, we find the following for the total SNR of the signal $\rho^2$
\begin{align}
\rho^2=\rho_g^2 + \rho_s^2 + \frac{A_gA_s}{5\pi^2 S}\left[\frac{16f_gf_sQ_g^3Q_s^3\left(f_gQ_s+f_sQ_g\right)\cos\phi}{\Lambda_+\Lambda_-}\right]\label{snrtotal}
\end{align}
with the individual SNRs for each mode being
\begin{align}
\rho_g^2=&\;\frac{A_g^2Q_g^3}{5\pi^2f_g\left(1+4Q_g^2\right)S}\\
\rho_s^2=&\;\frac{A_s^2Q_s\left(\sin^2\phi+2Q_s^2\right)}{10\pi^2f_s\left(1+4Q_s^2\right)S}
\end{align}
and where $\Lambda_\pm$ are given by
\begin{align}
\Lambda_\pm = f_s^2Q_g^2 + 2 f_g f_s Q_g Q_s + Q_s^2\left[f_g^2+4\left(f_g\pm f_s\right)^2Q_g^2\right].
\end{align}

This is the same result found in BCW for the total SNR of a two-mode waveform. BCW also showed that the phase $\phi$ only weakly affected their results, so for simplicity we chose to fix $\phi=\pi/2$ in the following so that the total SNR is simply given by the sum in quadrature of the individual SNRs. 

We now present analytic expressions for the error estimates for $A_g, M, A_s$ and $\mu^2$ (where $\sigma_a^2=\Sigma_{aa}$) to the \textit{leading order term} in the relative scalar amplitude $A_s/A_g$, as we assume that the amplitude of the scalar mode will be subdominant compared to the gravitational mode amplitude:  
\begin{subequations}
\begin{align}
\rho\frac{\sigma_{A_g}}{A_g}=&\;\sqrt{1+\frac{1}{Q_g^2}\frac{1+3Q_g^2}{3+8Q_g^2}}\nonumber\\
&\times\left[1+\left(\frac{A_s}{A_g}\right)^2\frac{\eta^2}{2}\left(1-\frac{\Lambda_2}{\Lambda_1}\frac{1+4Q_g^2}{1+2Q_g^2}\right)\right]\\
\rho\frac{\sigma_M}{M}=&\;\frac{1}{Q_g}\sqrt{\frac{1+4Q_g^2}{3+8Q_g^2}}\nonumber\\
&\times\left[1+\left(\frac{A_s}{A_g}\right)^2\frac{\eta^2}{2}\left(1-4\frac{\Lambda_2}{\Lambda_1}\left(1+4Q_g^2\right)\right)\right]\\
\rho\frac{\sigma_{A_s}}{A_s}=&\;\left(\frac{A_s}{A_g}\right)^{-1}\sqrt{\frac{\left(1+4Q_s^2\right)\Lambda_3}{\eta^2\Lambda_1}}\label{sigmaAs}\\
\rho\sigma_{\mu^2}=&\;f_sQ_s\left(\frac{A_s}{A_g}\right)^{-1}\sqrt{\frac{2\left(1+4Q_s^2\right)^3\left(1+2Q_s^2\right)}{\eta^2\Lambda_1}}.\label{sigmamu}
\end{align}
\end{subequations}
We've introduced $\eta^2$, $\Lambda_i$  to replace a number unwieldy expressions; the definitions of these terms can be found in appendix \ref{covarapp}. Note that the ratio of single waveform SNRs is given by:
\begin{align}
\frac{\rho_s}{\rho_g}=\eta\frac{A_s}{A_g}.
\end{align}
 
As expected, the error estimates of the non-scalar parameters become independent of any of the scalar waveform parameters as $A_s\to0$, with the leading order corrections entering at quadratic order in the scalar amplitude. For the scalar parameters, we see that the leading term for $\sigma_\mu^2$ scales as $(A_s/A_g)^{-1}$, thus diverging as the scalar amplitude $A_s\to0$ (as is reasonable).

Having calculated the error estimates analytically, we can consider the effect that introducing the scalar waveform has on $\sigma_{A_g}$ and $\sigma_M$. For $\ell=2$, $\mu^2=0$, the leading order corrections to the error estimates of $A_g$ and $M$ are given by:
\begin{align}
\frac{\rho\Delta\sigma_{A_g}}{A_g} \approx \; 0.52\left(\frac{A_s}{A_g}\right)^2 ,\;\;   \frac{\rho\Delta\sigma_M}{M}\approx &\; 0.13\left(\frac{A_s}{A_g}\right)^2
\end{align}
where again we are assuming that $A_s\ll A_g$. Clearly, with small $A_s$, the error estimates on $A_g$ and $M$ are only weakly degraded by the introduction of the scalar mode. We have also checked that the value of $(\mu M)^2$ only weakly affects $\sigma_M$ and $\sigma_A$. For $\sigma_{A_s}$ and $\sigma_{\mu^2}$, on the other hand, to leading order in both $A_s/A_g$ and $(\mu M)^2$ (again with $\ell=2$):
\begin{align}
\rho\sigma_{\mu^2}=&\;\frac{1}{M^2}\left(\frac{A_s}{A_g}\right)^{-1}\left(0.42-0.76\left(\mu M\right)^2\right)\\
\frac{\rho\sigma_{A_s}}{A_s}=&\;\left(\frac{A_s}{A_g}\right)^{-1}\left(1.00-0.40\left(\mu M\right)^2\right).
\end{align}

If we return to assuming that the effective scalar mass $\mu^2=0$, we can calculate $\sigma_{\mu^2}$ to estimate a `detectability' limit on the scalar `particle' effective mass $m_s$. Reinserting $G$ and $c$ to restore units, we find that
\begin{align}
\rho\sigma_{\mu^2}\sim 2\times10^{-7}\left(\frac{A_s}{A_g}\right)^{-1}\left(\frac{M_\odot}{M}\right)^2\text{m}^{-2},
\end{align}
where we now interpret $\mu^2$ as the square of the inverse Compton wavelength $\lambda_c$ of the scalar (thus the combination $\mu M$ is really a ratio of the black hole to scalar length scales). Converting to a mass using $m_s=h/\lambda_c c$, we find
\begin{align}
\sqrt{\rho} m_s \sim 5\times 10^{-10}\left(\frac{A_s}{A_g}\right)^{-1/2}\left(\frac{M_\odot}{M}\right)\text{eV}c^{-2}.\label{deltams}
\end{align}

\begin{figure}
\caption{Contour plot for $m_s$ as a function of relative scalar amplitude $A_s/A_g$ and black hole mass $M$ (in solar units) for values representative of LIGO events. The $[-8,...,-12]\times\log_{10}\left(\sqrt{\rho}m_s/\text{eV}c^{-2}\right)$ contours are shown.}
\label{LIGOerror}
\includegraphics[width=0.5\textwidth]{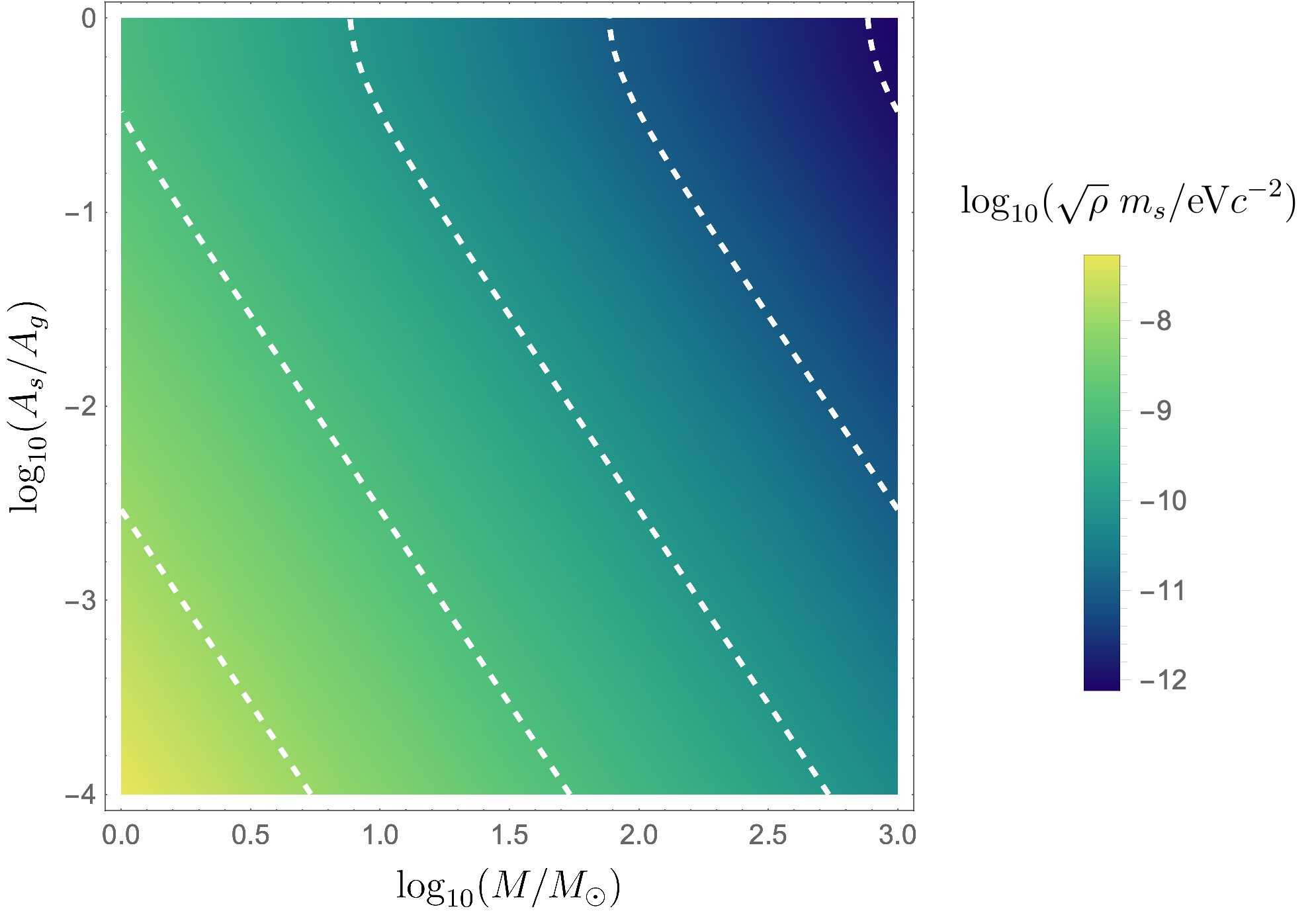}\\
\end{figure}

\begin{figure}
\caption{Contour plot for $m_s$ as a function of relative scalar amplitude $A_s/A_g$ and black hole mass $M$ (in solar units) for values representative of LISA events. The $[-13,...,-17]\times\log_{10}\left(\sqrt{\rho}m_s/\text{eV}c^{-2}\right)$ contours are shown.}
\label{LISAerror}
\includegraphics[width=0.5\textwidth]{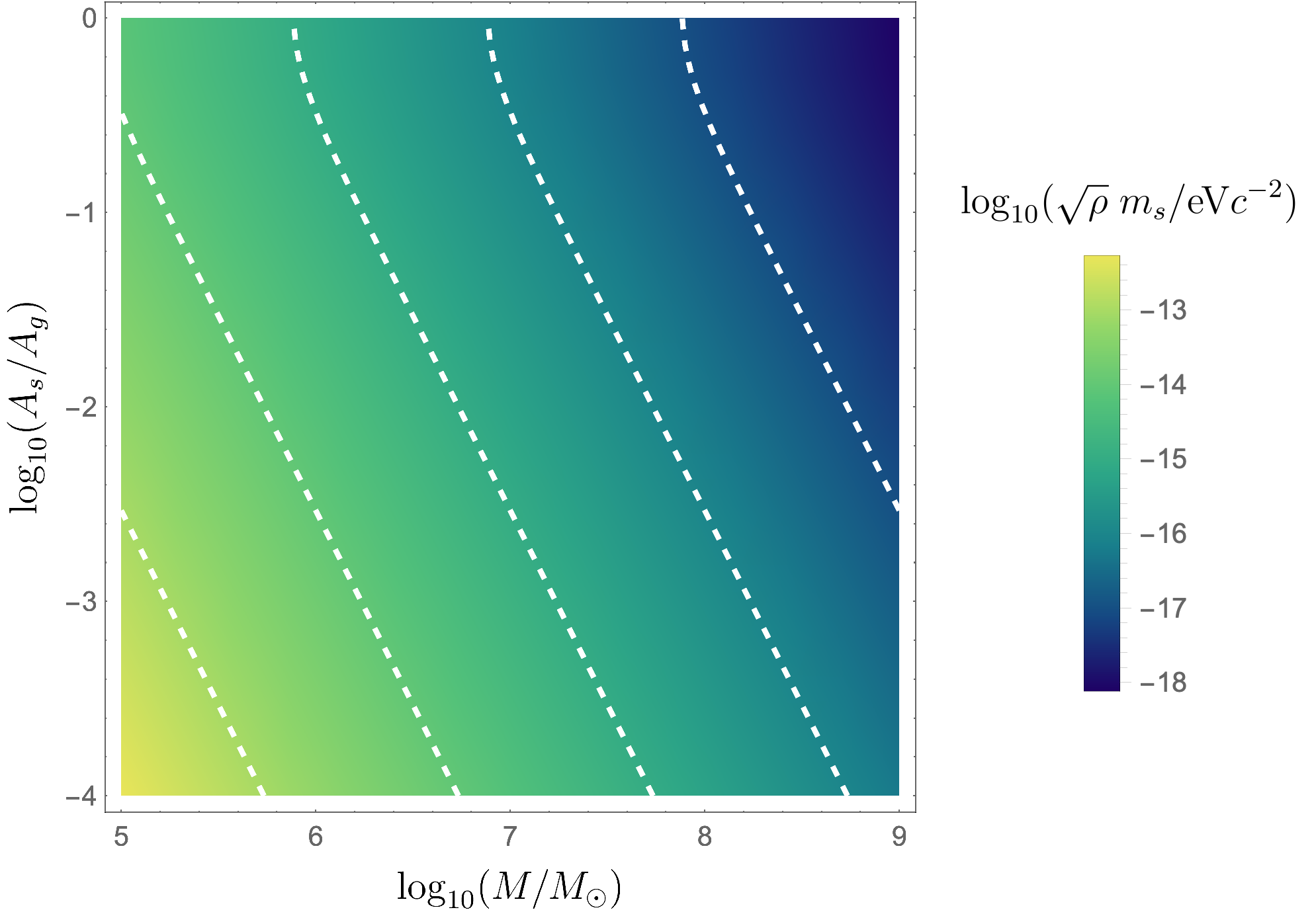}\\
\end{figure}

Figure \ref{LIGOerror} shows a contour plot of $\sqrt{\rho}m_s$ as a function of relative scalar amplitude $A_s/A_g$ and black hole mass $M$ for a mass range of likely events observed by aLIGO/VIRGO, while figure \ref{LISAerror} shows a similar contour plot but for more massive black holes of the type LISA might observe. In figure \ref{LIGOerror} a mass range of $0<\log_{10}(M/M_\odot)<3$ is covered, whilst for figure \ref{LISAerror} we consider a range of $5<\log_{10}(M/M_\odot)<9$, over which LISA is expected to be sensitive to BH ringdowns out to large redshifts \cite{Baibhav:2018rfk}. Note that in figures $\ref{LIGOerror}$ and $\ref{LISAerror}$ we have used the full expressions for the covariance matrix as calculated using the Fisher matrix formalism, rather than the $A_s\ll A_g$ approximation used to arrive at eq.~(\ref{deltams}).

In figures \ref{LIGOerror} and \ref{LISAerror} we see that with increasing black hole mass the constraint on $m_s$ tightens considerably, even for small $A_s/A_g$. For example, constraints on $\sqrt{\rho}m_s\sim10^{-17}\text{eV}c^{-2}$ are possible with $A_s/A\sim0.01$ for a $10^9M_\odot$ black hole. 


\subsubsection{$j\neq0$}

We now consider the case of a slowly rotating black hole with $j\ll1$. In this case the total SNR is still given by eq.~(\ref{snrtotal}), and again we choose $\phi=\pi/2$. The introduction of $j$ into the Fisher matrix analysis makes the analytic expressions for the error estimates extremely unwieldy, so we will not present the leading order corrections due to $A_s$ to $\sigma_{A_g}$, $\sigma_{j}$ or $\sigma_M$. Instead we direct the reader to section VI B of BCW, which demonstrates the effect of increasing black hole spin on the error estimates of $j$ and $M$ in a two-mode waveform; we will focus on the effect of $j$ on $\sigma_{A_s}$ and $\sigma_\mu^2$.

In fact, we find that to leading order in $(A_s/A_g)$, that the error estimates of $A_s$ and $\mu^2$ are given once again by eq.~(\ref{sigmaAs})-(\ref{sigmamu}). We can now evaluate these numerically for non-zero (small) $j$. For $\ell=2$, $\mu^2=0$, we find the following error estimates to linear order in $j$ and to leading order in $A_s/A_g$:
\begin{align}
\rho \frac{\sigma_{A_s}}{A_s}=&\;\left(1.00+0.03jm\right)\left(\frac{A_s}{A_g}\right)^{-1}\\
\rho \sigma_{\mu^2}=&\;\frac{0.42}{M^2}\left(1+\frac{jm}{3}\right)\left(\frac{A_s}{A_g}\right)^{-1},
\end{align}
where $m$ is the azimuthal spherical harmonic index ranging from $(-\ell,...,\ell)$. This corresponds to a scalar mass detectability limit of 
\begin{align}
\sqrt{\rho}m_s \sim 5\times 10^{-10}(1+\frac{jm}{3})\left(\frac{A_s}{A_g}\right)^{-1/2}\left(\frac{M_\odot}{M}\right)\text{eV}c^{-2},
\end{align}
again valid for $j\ll1$, $A_s\ll A_g$. We see that, in this slow rotation regime, the introduction of spin weakly increases (decreases) the error estimates for the scalar parameters for positive (negative) $m$. 

It would of course be interesting to repeat this analysis for larger values of $j$, however we are currently unaware of any fitting formulae for massive scalar QNMs as a function of both scalar mass \textit{and} black hole spin. 

The phenomenon of black hole superradiance \cite{Brito:2015oca} provides a method of constraining ultralight boson masses through observations of rotating black holes \cite{Arvanitaki:2016qwi,Brito:2017wnc,Brito:2017zvb,Cardoso:2018tly,Berti:2019wnn}. In \cite{Cardoso:2018tly} masses of minimally coupled axion like particles are excluded in the range $[6\times 10^{-13}\text{eV},10^{-11}\text{eV}]$, and it is argued that current and future observations can probe a mass range of such particles from $10^{-19}$eV to $10^{-11}$eV. In fact, if we posit that a modified gravity theory admits Kerr black hole solutions (and that these are indeed the astrophysical black holes we observe), then the same bounds calculated from superradiant instabilities in \cite{Cardoso:2018tly} apply equally to such modified gravity theories.

Whilst it would take a very large black hole mass $M$ or SNR $\rho$ to compete with the lowest end of the mass range probed by back hole superradiance, the constraints that could be garnered from ringdown observations can be complementary to those obtained from other methods. Furthermore, in this case we are considering a phenomenon arising specifically from the non-minimal coupling between gravity and the scalar field.

\subsection{Resolvability}

To use a mixed-mode GW signal to test GR one must, of course, first be able to discriminate between the two frequencies buried in the noisy signal. As explored in BCW as well as in \cite{Berti:2007zu,Molina:2010fb}, such considerations lead to postulating a minimum SNR required to resolve the individual frequencies and damping times of the gravitational and scalar modes. It is commonly given that a natural criterion for resolving frequencies and damping times is given by
\begin{align}
|f_g-f_s|>\text{max}\left(\sigma_{f_g},\sigma_{f_s}\right),\;|\tau_g-\tau_s|>\text{max}\left(\sigma_{\tau_g},\sigma_{\tau_s}\right).
\end{align}
As explained in BCW, the above criteria states that the frequencies are barely resolvable if ``the maximum of the diffraction pattern of object 1 is located at the minimum of the diffraction pattern of object 2''. Using the above, critical SNRs required to resolve the individual frequencies and damping times can be introduced
\begin{align}
\rho>\rho_{\text{crit}}^{f}=&\;\frac{\text{max}\left(\rho\sigma_{f_g},\rho\sigma_{f_s}\right)}{|f_g-f_s|}\\
\rho>\rho_{\text{crit}}^{\tau}=&\;\frac{\text{max}\left(\rho\sigma_{\tau_g},\rho\sigma_{\tau_s}\right)}{|\tau_g-\tau_s|}.\label{rhocritt}
\end{align}

We can use the Fisher matrix formalism to calculate these error estimates analytically, now in the $(f,\tau)$ parameter basis. Again to leading order in $A_s/A_g$, we find the following expressions for the errors in $f$ and $\tau$ for each waveform:
\begin{subequations}
\begin{align}
\rho \sigma_{f_g}=&\;\frac{f_g}{8Q_g^3}\sqrt{6+32Q_g^4}\left(1+\left(\frac{A_s}{A_g}\right)^2\frac{\eta^2}{2}\right)\\\label{rhocritfg}
\rho \sigma_{\tau_g}=&\;\frac{1}{\pi f_g}\sqrt{3+4Q_g^2}\left(1+\left(\frac{A_s}{A_g}\right)^2\frac{\eta^2}{2}\right)\\
\rho \sigma_{f_s}=&\;\frac{f_s}{2\sqrt{2}Q_s^2 \eta}\left(\frac{A_s}{A_g}\right)^{-1}\sqrt{\frac{1+4Q_s^4\left(7+16Q_s^2\left(2+Q_s^2\right)\right)}{1+2Q_s^4\left(5+8Q_s^2\left(2+Q_s^2\right)\right)}}\\
\rho \sigma_{\tau_s}=&\;\frac{\sqrt{2}Q_s}{\pi f_s \eta}\left(\frac{A_s}{A_g}\right)^{-1}\sqrt{1+\frac{3+4Q_s^2-24Q_s^4}{1+2Q_s^4\left(5+8Q_s^2\left(2+Q_s^2\right)\right)}}\label{rhocritts}
\end{align}
\end{subequations}
with $\eta^2$ given in appendix \ref{covarapp}.

Figure \ref{rhocritplot} shows the two critical SNRs $\rho_{\text{crit}}^f$ and $\rho_{\text{crit}}^\tau$ required to resolve individual frequencies and damping times respectively as a function of the relative scalar mode amplitude. We've again chosen a superposition of $\ell=2$ modes for a Schwarzschild black hole and a massless scalar field. As with figures \ref{LIGOerror} and \ref{LISAerror}, here we have used the full analytic expressions for the critical SNRs in figure \ref{rhocritplot}, rather than the $A_s\ll A_g$ approximation used in eqs.~(\ref{rhocritfg})-(\ref{rhocritts}). 

We see that the SNR required to resolve damping times is consistently about an order of magnitude greater than that required to resolve frequency times. Thus $\rho_{\text{crit}}^\tau$ sets the lower bound on SNR to resolve \textit{both} frequencies and damping times. The minimum of $\rho_{\text{crit}}^\tau$ is found numerically, in this case, to be at $A_s/A_g\approx0.81$, giving a critical SNR of 33.5. 

Assuming that we wish to resolve both frequencies and damping times, and noting that for small $A_s$ the critical SNR is given by $\rho \sigma_{\tau_s}$, we can use eq.~(\ref{rhocritt}) and (\ref{rhocritts}) to find a minimum requirement on $A_s/A_g$ for a given SNR. For a superposition of $\ell=2$ modes with $\mu^2=0$, we find
\begin{align}
\frac{A_s}{A_g} > \frac{21}{\rho}.
\end{align}
If we only wish to distinguish frequencies and not damping times, the requirement on $A_s$ drops to
\begin{align}
\frac{A_s}{A_g}>\frac{1.2}{\rho}.
\end{align}

For example, with an SNR of $\rho\sim10^2$ (achievable in single detections through LISA, third generation ground based detectors, or through stacking several signals together \cite{Berti:2016lat,Yang:2017zxs,Yang:2017xlf,DaSilvaCosta:2017njq,2018PhRvD..98h4038B,2019arXiv190209199B}), we would require $A_s\approx 0.2$ to ensure $\rho>\rho_{\text{crit}}^{\tau}$. If, on the other hand, we considered single, loud, aLIGO/VIRGO detections such as GW150914, an SNR of $\rho\sim5-10$ is more realistic \cite{2016PhRvL.116v1101A}. In which case $A_s\approx0.2$ would be required just to discern distinct oscillation frequencies in the signal, whilst an observation of distinct damping times would be impossible given that the minimum of $\rho_{\text{crit}}^\tau$ is $33.5$ as discussed above (and as shown in figure \ref{rhocritplot}).

\begin{figure}
\caption{Critical SNRs $\rho_{\text{crit}}^f$ and $\rho_{\text{crit}}^\tau$ required to resolve frequencies and damping times as a function of relative scalar amplitude $A_s/A$ for $\ell=2$, $\mu^2=j=0$.}
\label{rhocritplot}
\includegraphics[width=0.49\textwidth]{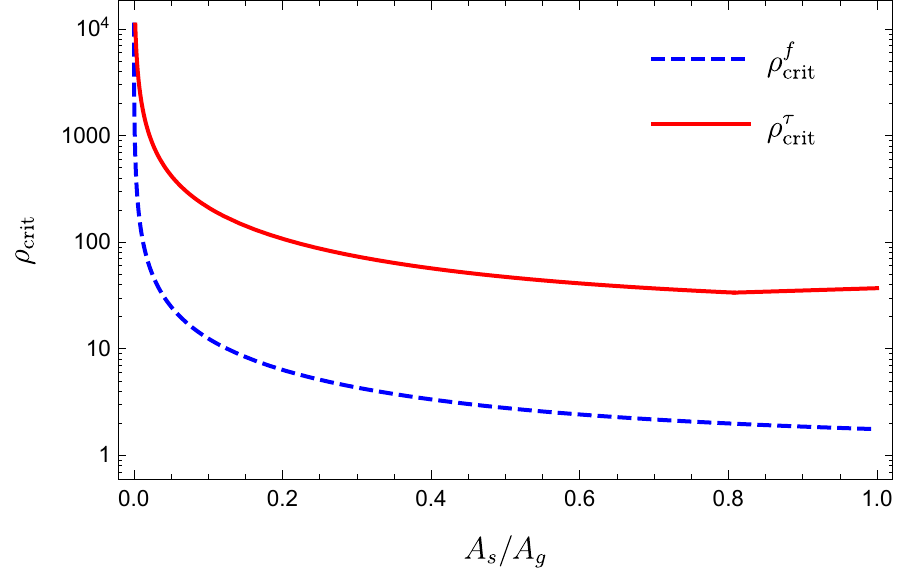}\\
\end{figure}

\section{Discussion \& limitations}\label{discussion}
In this work we have revisited the mixing of gravitational and scalar modes in the GW emission during ringdown of static and slowly rotating black holes in Horndeski gravity as first described in \cite{2018PhRvD..97d4021T,Tattersall:2018nve}. The qualitative nature of the mode mixing effect on the GW emission is shown in figure \ref{cartoon}, with a related phenomenon in scalar Gauss-Bonnet gravity has been observed numerically in \cite{Witek:2018dmd}. Additionally, we demonstrated its occurence in any Ricci flat black hole background (see eq.~(\ref{einseq})). Indeed, a natural progression of this work is to study in detail the perturbations of Kerr black holes of arbitrary spin in Horndeski gravity, through both analytical and numerical methods. 

We then proceeded to apply the Fisher matrix formalism for black hole ringdown as developed in BCW \cite{Berti:2005ys} to a mixed mode waveform containing both gravitational and scalar frequencies, and in section \ref{secresults} derived analytic expressions for the estimated errors on the parameters of such a waveform assuming a static or slowly rotating black hole background. Of particular interest is the estimated error in the determination of the effective mass of the Horndeski scalar field (see eq.~(\ref{sigmamu}), and figures \ref{LIGOerror} and \ref{LISAerror}). For certain parameter ranges of the black hole mass and relative scalar mode amplitude, we've shown that constraints on the effective mass of the Horndeski scalar field can be very tight, for example $m_s \sim \rho^{-1/2}10^{-17}$eV$c^{-2}$ for a $10^9M_\odot$ black hole observed with LISA; competitive with the kind of constraints on ultralight axion masses obtained via black hole superradiance. 

We further found that, assuming an SNR of $\rho\sim 10^2$  (achievable through next generation space and ground based detectors or through the stacking of multiple signals \cite{Berti:2016lat,Yang:2017zxs,Yang:2017xlf,DaSilvaCosta:2017njq,2018PhRvD..98h4038B,2019arXiv190209199B}) a scalar perturbation with an amplitude of roughly $20$\% that of the dominant gravitational mode would be required so that the presence of multiple, distinct oscillations frequencies and damping times in the signal could be detected. With the SNRs typical of single LIGO events \cite{2016PhRvL.116v1101A}, on the other hand, detecting the presence of distinct oscillation frequencies in the ringdown signal is the best one can hope for, with a scalar mode amplitude again of the order $10-20$\% that of the gravitational mode amplitude required. 

A key assumption in this work is that the scalar perturbations will be present in the ringdown: specifically, if  mode mixing is to occur in the ringdown, the scalar field perturbation needs to be excited, which is by no means guaranteed given that $\varphi=0$ is a solution to eq.~(\ref{RWeq})-(\ref{Phieq}) (leading to perturbations identical to those in GR). Furthermore, in the case of two black holes without hair merging, it is difficult to envisage generating any scalar field perturbations (though perhaps non-linear interactions during the merger could source excitations). However, if $\phi$ interacts non-trivially with matter,  then events involving one or more compact stars may provide an initial non-trivial scalar field profile from which perturbations can be sourced \cite{Silva:2014fca,Pani:2014jra,Maselli:2016gxk,Babichev:2016rlq,Minamitsuji:2016hkk}. In addition, one could imagine merger events whereby surrounding matter `contaminates' the `clean' merger of two compact objects without hair, thus sourcing scalar perturbations in situations where one might not initially expect them. An important avenue of research is then to study the possibility of how hair can be dynamically generated in no-hair theories.

An obvious limitation of this work is the inclusion of only two modes in the mixed-mode waveform - one from each of the gravitational and scalar spectra. In \cite{Giesler:2019uxc} it is shown that including higher overtones as well as fundamental modes is highly important for accurately extracting parameters from a GW signal. In addition, the study of mode amplitudes in black hole ringdown \cite{Berti:2006wq,Kamaretsos:2011um,Zhang:2013ksa,London:2014cma} shows that in some cases the second most dominant gravitational mode may have a significant relative amplitude of $\mathcal{O}(1)$. In which case modelling our signal as a two-mode waveform with $A_2 \ll A_1$ may be a simplification too far (of course if a scenario heavily excited the scalar mode then the two mode approach would be more valid). Including additional gravitational modes in addition to the  scalar mode(s) would be a more accurate approach, and an intriguing area of future research. 

Finally, the numerical results shown in this paper are limited to Schwarzschild and slowly rotating Kerr black holes, but of course it is our aim to apply such an analysis to Kerr black holes of arbitrary spin (especially given the so far observed spins of black holes by aLIGO/VIRGO). Analytic fits of massive scalar QNM frequencies on a Kerr background, or a more in depth numerical analysis, will be required to make such a step to higher spins; these are interesting areas of future work.  

With the maturation of GW astronomy, and the prospect of black hole spectroscopy with next generation detectors in the near future, the exploration of what we can learn about the nature of gravity from ringdown observations is an exciting and timely endeavour.

%\section{Conclusion}\label{conclusion}

\section*{Acknowledgments}
\vspace{-0.2in}
\noindent We're grateful to E. Berti, V. Cardoso, and K. Clough for useful discussions and comments throughout the preparation of this work. OJT was supported by the Science and Technology Facilities Council (STFC) Project Reference 1804725. PGF acknowledges support from STFC, the Beecroft Trust and the European Research Council.

\appendix

\section{Source Pattern Functions}\label{fisherapp}

The source pattern functions $F_+$ and $F_\times$ referred to in eq.~(\ref{basich}) are given by \cite{Berti:2005ys}:
\begin{subequations}
\begin{align}
F_+=&\;\frac{1}{2}\left(1+\cos^2\theta_S\right)\cos2\phi_S\cos2\psi_S-\cos\phi_S\sin2\phi_S\sin2\psi_S\\
F_\times=&\;\frac{1}{2}\left(1+\cos^2\theta_S\right)\cos2\phi_S\sin2\psi_S+\cos\phi_S\sin2\phi_S\cos2\psi_S.
\end{align}
\end{subequations}
The angles $\theta_S$ and $\phi_S$ give the angular position of the GW source in usual spherical coordinates, whilst $\psi_S$ describes the rotation of the GW polarisation axes relative to the detector arm axes \cite{Schutz:2011tw}. 

\section{Error Estimate Expressions}\label{covarapp}

The terms $\eta^2$ and $\Lambda_i$ were introduced in section \ref{secresults} for brevity's sake. Their explicit expressions are given by:
\begin{subequations}
\begin{align}
\eta^2=&\;\frac{1}{2}\frac{f_gQ_s}{f_sQ_g}\frac{1+4Q_g^2}{Q_g^2}\frac{1+2Q_s^2}{1+4Q_s^2}\label{etasq}\\
\Lambda_1=&\;f_s^2 \left(4 \left(16 \left(Q_s^2+2\right) Q_s^2+7\right)
   Q_s^4+1\right) Q_{s,\mu^2}^2\nonumber\\
   &-2 f_s Q_s f_{s,\mu^2} Q_{s,\mu^2} \left(64
   Q_s^8+64 Q_s^6+28 Q_s^4+8 Q_s^2+1\right)\nonumber\\
   &+f_{s,\mu^2}^2 Q_s^2 \left(32 Q_s^6+28 Q_s^4+8 Q_s^2+1\right)
   \left(1+4Q_s^2\right)^2\\
   \Lambda_2=&\;\frac{2 M^2 Q_s^4 \left(16 Q_s^6+32 Q_s^4+10 Q_s^2+1\right)}{Q^2 \left(8 Q^2+3\right)}\nonumber\\
   &\times (f_{s,\mu^2} Q_{s,M}-f_{s,M} Q_{s,\mu^2})^2\\
   \Lambda_3=&\;f_s^2 \left(32 Q_s^6+24 Q_s^4+1\right) Q_{s,\mu^2}^2\nonumber\\
   &-2 f_s
   f_{s,\mu^2} \left(32 Q_s^7+16 Q_s^5+6 Q_s^3+Q_s\right)
   Q_{s,\mu^2}\nonumber\\
   &+f_{s,\mu^2}^2 Q_s^2\left(8 Q_s^4+4 Q_s^2+1\right) \left(1+4Q_s^2\right)^2,
\end{align}
\end{subequations}
where we are using the notation $\mathcal{F}_{,\theta}=\frac{\partial \mathcal{F}}{\partial \theta}$.

\bibliography{RefModifiedGravity}

\end{document}